\documentclass[twocolumn,eqsecnum,prb]{revtex4}
\usepackage{amsfonts}
\usepackage[T1]{fontenc}
\usepackage{amsmath,amsbsy,amssymb,graphicx}
\usepackage{times}
\usepackage{amsmath}
\usepackage{amssymb}
\usepackage{graphicx}

\let\mathbf=\boldsymbol
\def\beginABC{\begin{subequations}}
\def\endABC{\end{subequations}}
\begin{document}

\title{{\Large Topological Insulator and Helical Zero Mode in Silicene }\\
{\Large under Inhomogeneous Electric Field}}
\author{Motohiko Ezawa}
\affiliation{Department of Applied Physics, University of Tokyo, Hongo 7-3-1, 113-8656,
Japan }

\begin{abstract}
Silicene is a monolayer of silicon atoms forming a two-dimensional honeycomb
lattice, which shares almost every remarkable property with graphene. The
low energy structure of silicene is described by Dirac electrons with
relatively large spin-orbit interactions due to its buckled structure. The
key observation is that the band structure is controllable by applying the
electric field to a silicene sheet. In particular, the gap closes at a
certain critical electric field. Examining the band structure of a silicene
nanoribbon, we demonstrate that a topological phase transition occurs from a
topological insulator to a band insulator with the increase of the electric
field. We also show that it is possible to generate helical zero modes
anywhere in a silicene sheet by adjusting the electric field locally to this
critical value. The region may act as a quantum wire or a quantum dot
surrounded by topological and/or band insulators. We explicitly construct
the wave functions for some simple geometries based on the low-energy
effective Dirac theory. These results are applicable also to germanene, that
is a two-dimensional honeycomb structure of germanium.
\end{abstract}

\maketitle

%\date{}

\address{{\normalsize Department of Applied Physics, University of Tokyo, Hongo
7-3-1, 113-8656, Japan }}

\section{Introduction}

Graphene, a monolayer honeycomb structure of carbon atoms, is one of the
most important topics in condensed matter physics\cite{GrapheneRMP}. One of
the obstacles of graphene for electronic devices is that electrons can not
be confined by applying external electric field\cite{Klein}. Thus, graphene
nanostructures such as graphene nanoribbon\cite{Nanoribbon} and nanodisk\cite%
{Nanodisk} have been considered, which are to be fabricated by cutting a
graphene sheet. Recently a new material, a monolayer honeycomb structure of
silicon called silicene, has been synthesized\cite{Lalmi,Padova,Aufray} and
attracts much attention\cite{Guzman,LiuPRL,LiuPRB}. Silicene has Dirac cones
akin to graphene. Almost every striking property of graphene could be
transferred to this innovative material. Furthermore, silicene has advantage
of easily being incorporated into the silicon-based electronic technology.

\begin{figure}[t]
\centerline{\includegraphics[width=0.45\textwidth]{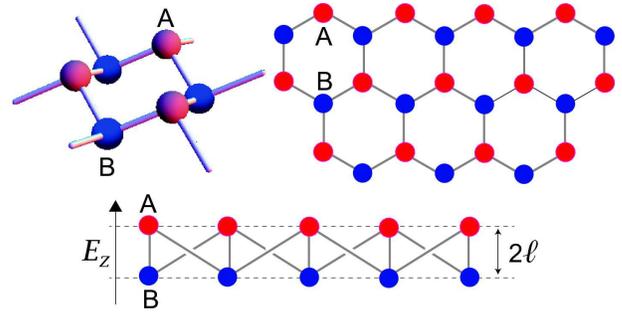}}
\caption{(Color online) Illustration of the buckled honycomb lattice of
silicene. A honeycomb lattice is distorted due to a large ionic radius of a
silicon atom and forms a buckled structure. The A and B sites form two
sublattices separated by a perpendicular distance $2\ell$. The structure
generates a staggered sublattice potential in the electric field $E_{z}$,
which leads to various intriguing pheneomena.}
\label{FigBuckl}
\end{figure}

Silicene has a remarkable property graphene does not share: It is the
buckled structure\cite{LiuPRL,LiuPRB} owing to a large ionic radius of
silicon (Fig.\ref{FigBuckl}). Consequently, silicene has a relatively large
spin-orbit (SO) gap of $1.55$meV, as makes experimentally accessible the
Kane-Mele type quantum Spin Hall (QSH) effect or topological insulator\cite%
{LiuPRL,LiuPRB}. Topological insulator\cite{Hasan,Qi} is a new state of
quantum matter characterized by a full insulating gap in the bulk and
gapless edges topologically protected. These states are made possible due to
the combination of the SO interaction and the time-reversal symmetry. The
two-dimensional topological insulator is a QSH insulator with helical
gapless edge modes\cite{Wu}, which is a close cousin of the integer quantum
Hall state. QSH insulator was proposed by Kane and Mele in graphene\cite%
{KaneMele}. However, since the SO gap is rather weak in graphene, the QSH
effect can occur in graphene only at unrealistically low temperature\cite%
{Min, Yao}.

The buckled structure implies an intriguing possibility that we can control
the band structure by applying the electric field (Fig.\ref{FigBuckl}). In
this paper, we analyze the band structure under the electric field $E_{z}$
applied perpendicular to a silicene sheet. Silicene is a $\mathbb{Z}_{2}$\
topological insulator\cite{LiuPRL} at $E_{z}=0$. By increasing $E_{z}$, we
demonstrate the following. The gap decreases linearly to zero at a certain
critical field $E_{c}$ and then increases linearly. Accordingly, silicene
undergoes a topological phase transition from a topological insulator to a
band insulator. At the critical point ($E_{z}=E_{c}$), spins are perfectly
spin-up (spin-down) polarized at the K (K') point.

We also investigate the zero-energy states under an inhomogeneous electric
field $E_{z}(x,y)$ based on the low-energy effective Dirac theory. There
emerge helical zero modes in the region where $E_{z}(x,y)=E_{c}$. It is
intriguing that the region is not necessary one-dimensional: The region can
be two-dimensional and have any shape, where Dirac electrons can be
confined. It is surrounded by topological and/or band insulators. Our result
may be the first example in which helical zero modes appear in regions
besides the edge of a topological insulator. The region may act as a quantum
wire or a quantum dot. We construct explicitly the wave functions describing
helical zero modes for regions having simple geometries. In conclusion, we
are able to realize a dissipationless spin current anywhere in the bulk of a
silicene sheet by tuning the electric field locally.

\section{Topological and Band insulators}

Silicene consists of a honeycomb lattice of silicon atoms with two
sublattices made of A sites and B sites. The states near the Fermi energy
are $\pi $ orbitals residing near the K and K' points at opposite corners of
the hexagonal Brillouin zone. We take a silicene sheet on the $xy$-plane,
and apply the electric field $E_{z}(x,y)$ perpendicular to the plane. Due to
the buckled structure the two sublattice planes are separated by a distance,
which we denote by $2\ell $ with $\ell =0.23$\AA\ , as illustrated in Fig.%
\ref{FigBuckl}. It generates a staggered sublattice potential $\varpropto
2\ell E_{z}(x,y)$ between silicon atoms at A sites and B sites.

\begin{figure}[t]
\centerline{\includegraphics[width=0.27\textwidth]{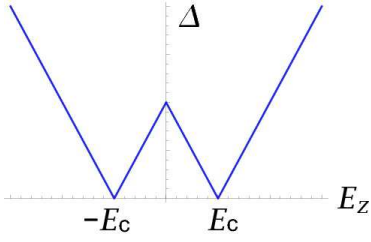}}
\caption{(Color online) The band gap $\Delta $ as a function of the electric
field $E_{z}$. The gap is open for $E_{z}\neq \pm E_{c}$, where silicene is
an insulator. It can be shown that it is a topological insulator for $%
|E_{z}|<E_{c}$ and a band insulator $|E_{z}|>E_{c}$.}
\label{FigGap}
\end{figure}

The silicene system is described by the four-band second-nearest-neighbor
tight binding model\cite{LiuPRB},

\begin{align}
H& =-t\sum_{\left\langle i,j\right\rangle \alpha }c_{i\alpha }^{\dagger
}c_{j\alpha }+i\frac{\lambda _{\text{SO}}}{3\sqrt{3}}\sum_{\left\langle
\left\langle i,j\right\rangle \right\rangle \alpha \beta }\nu
_{ij}c_{i\alpha }^{\dagger }\sigma _{\alpha \beta }^{z}c_{j\beta }  \notag \\
& -i\frac{2}{3}\lambda _{\text{R}}\sum_{\left\langle \!\left\langle
i,j\right\rangle \!\right\rangle \alpha \beta }\mu _{ij}c_{i\alpha
}^{\dagger }\left( \vec{\sigma}\times \vec{d}_{ij}^{0}\right) _{\alpha \beta
}^{z}c_{j\beta }  \notag \\
& +\ell \sum_{i\alpha }\zeta _{i}E_{z}^{i}c_{i\alpha }^{\dagger }c_{i\alpha
}.  \label{BasicHamil}
\end{align}%
The first term represents the usual nearest-neighbor hopping on the
honeycomb lattice with the transfer energy $t=1.6$eV, where the sum is taken
over all pairs $\left\langle i,j\right\rangle $ of the nearest-neighboring
sites, and the operator $c_{i\alpha }^{\dagger }$ creates an electron with
spin polarization $\alpha $ at site $i$. The second term represents the
effective SO coupling with $\lambda _{\text{SO}}=3.9$meV, where $\vec{\sigma}%
=(\sigma ^{x},\sigma ^{y},\sigma ^{z})$ is the Pauli matrix of spin, $\nu
_{ij}=\left( \vec{d}_{i}\times \vec{d}_{j}\right) /\left\vert \vec{d}%
_{i}\times \vec{d}_{j}\right\vert $ with $\vec{d}_{i}$ and $\vec{d}_{j}$ the
two nearest bonds connecting the next-nearest neighbors, and the sum is
taken over all pairs $\left\langle \!\left\langle i,j\right\rangle
\!\right\rangle $ of the second-nearest-neighboring sites. The third term
represents the Rashba SO coupling with $\lambda _{\text{R}}=0.7$meV, where $%
\mu _{ij}=\pm 1$ for the A (B) site, and $\vec{d}_{ij}^{0}=\vec{d}%
_{ij}/\left\vert \vec{d}_{ij}\right\vert $. The forth term is the staggered
sublattice potential term, where $\zeta _{i}=\pm 1$ for the A (B) site. Note
that the first and the second terms constitute the Kane-Mele model proposed
to demonstrate the QSH effect in graphene\cite{KaneMele}.

The same Hamiltonian as (\ref{BasicHamil}) can be used to describe
germanene, that is a honeycomb structure of germanium\cite{LiuPRL,LiuPRB},
where various parameters are $t=1.3$eV, $\lambda _{\text{SO}}=43$meV, $%
\lambda _{\text{R}}=10.7$meV and $\ell =0.33$\AA . Hence the following
analysis is applicable to germanene as well.

We study the band structure of silicene by applying a uniform electric field 
$E_{z}$. By diagonalizing the Hamiltonian (\ref{BasicHamil}), the band gap $%
\Delta\left( E_{z}\right) $ is determined to be%
\begin{equation}
\Delta\left( E_{z}\right) =2\left\vert \ell E_{z}-\eta s_{z}\lambda _{\text{%
SO}}\right\vert ,  \label{gap}
\end{equation}
where $s_{z}=\pm1$ is the electron spin and $\eta=\pm1$ is for the K or K'
point (to which we refer also as the K$_{\pm}$ point). See also the
dispersion relation (\ref{gapDirac}) which we derive based on the low-energy
effective theory. We emphasize that it is independent of the Rashba SO
coupling $\lambda_{\text{R}}$. The gap (\ref{gap}) vanishes at $E_{z}=\eta
s_{z}E_{c}$ with%
\begin{equation}
E_{c}=\lambda_{\text{SO}}/\ell=17\text{meV/\AA }.  \label{CritiE}
\end{equation}
We plot the band gap $\Delta\left( E_{z}\right) $ in the Fig.\ref{FigGap}.

\begin{figure}[t]
\centerline{\includegraphics[width=0.4\textwidth]{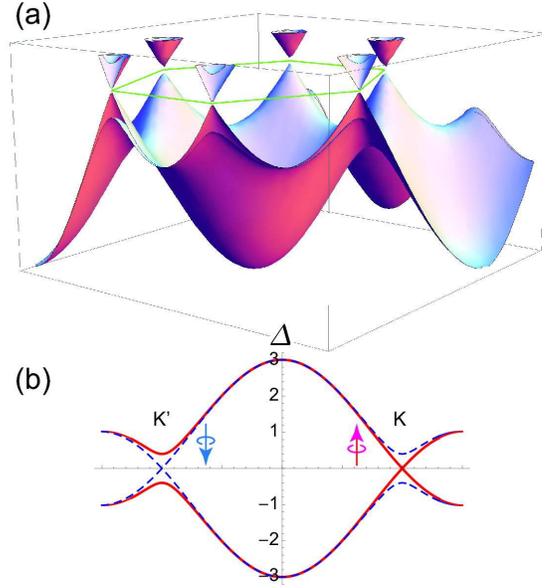}}
\caption{(Color online) Band structure of silicene at the critical electric
field $E_{c}$. (a) A bird's-eye view. Dirac cones are found at 6 corners of
the hexagonal Brillouin zone. (b) The cross section containing a pair of K
and K' points. The solid red (dashed blue) band is for up-spin (down-spin)
electrons, which are gapless (gapped) at the K point but gapped (gapless) at
the K' point.}
\label{FigBand}
\end{figure}

The gap closes at $E_{z}=\pm E_{c}$, where it is a semimetal due to gapless
modes. We show the band structure at $E_{z}=E_{c}$ in Fig.\ref{FigBand}. It
follows from (\ref{gap}) that up-spin ($s_{z}=+1$) electrons are gapless at
the K point ($\eta =+1$), while down-spin ($s_{z}=-1$) electrons are gapless
at the K' point ($\eta =-1$). Namely, spins are perfectly up (down)
polarized at the K (K') point under the uniform electric field $E_{z}=E_{c}$.

\begin{figure}[t]
\centerline{\includegraphics[width=0.5\textwidth]{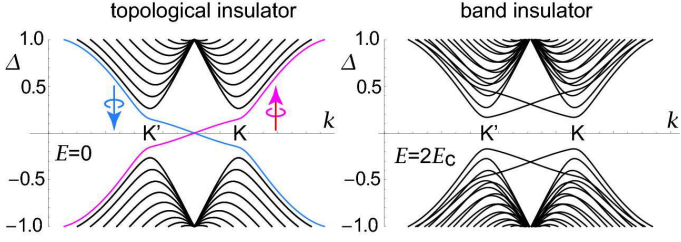}}
\caption{(Color online) One-dimensional energy bands for a silicene
nanoribbon. (a) The bands crossing the gap are edge states, demonstrating
that it is a topological insulator. There are two edge states since a
nanoribbon has two edges (red and blue lines for the left and right edges).
(b) All states are gapped, demonstrating that it is a band insulator.}
\label{FigSRibbonBand}
\end{figure}

It follows from the gap formula (\ref{gap}) that silicene is an insulator
for $E_{z}\neq \pm E_{c}$. In order to tell the difference between the two
insulators realized for $\left\vert E_{z}\right\vert <E_{c}$ and $\left\vert
E_{z}\right\vert >E_{c}$, we study the band structure of a silicene
nanoribbon with zigzag edges. The gap structure is depicted at two typical
points, $E_{z}=0$ and $E_{z}=2E_{c}$, in Fig.\ref{FigSRibbonBand}. We see
that there are gapless modes coming from the two edges at $\left\vert
E_{z}\right\vert <E_{c}$, as is the demonstration of a topological insulator%
\cite{LiuPRL}. On the other hand, there are no gapless edge modes for $%
\left\vert E_{z}\right\vert >E_{c}$, showing that it is a band insulator. We
conclude that a topological phase transition occurs between a topological
insulator ($\left\vert E_{z}\right\vert <E_{c}$) and a band insulator ($%
|E_{z}|>E_{c}$) as $E_{z}$ changes.

The reason why gapless modes appear in the edge of a topological insulator
is understood as follows. The topological insulator has a nontrivial
topological number, the $\mathbb{Z}_{2}$ index\cite{KaneMele}, which is
defined only for a gapped state. When a topological insulator has an edge
beyond which the region has the trivial $\mathbb{Z}_{2}$ index, the band
must close and yield gapless modes in the interface. Otherwise the $\mathbb{Z%
}_{2}$ index cannot change its value across the interface.

\section{Low-Energy Dirac Theory}

We proceed to analyze the physics of electrons near the Fermi energy more in
details. The low-energy Dirac theory has been proved to be essential in the
study of graphene\cite{Semenoff} and its various derivatives\cite%
{Brey73,EzawaDirac}. It must also be indispensable to explore deeper physics
of helical zero modes and promote further researches in silicene.

We may derive the low-energy effective Hamiltonian from the tight binding
model (\ref{BasicHamil}) around the $K_{\eta}$ point as\cite{LiuPRB}%
\begin{equation}
H_{\eta}=\hbar v_{\text{F}}\left( k_{x}\tau_{x}-\eta k_{y}\tau_{y}\right)
+\eta\tau_{z}h_{11}+\ell E_{z}\tau_{z},  \label{DiracHamil}
\end{equation}
with%
\begin{equation}
h_{11}=-\lambda_{\text{SO}}\sigma_{z}-a\lambda_{\text{R}}\left(
k_{y}\sigma_{x}-k_{x}\sigma_{y}\right) ,
\end{equation}
where $\tau_{a}$ is the Pauli matrix of the sublattice, $v_{\text{F}}=\frac{%
\sqrt{3}}{2}at=5.5\times10^{5}$m/s is the Fermi velocity, and $a=3.86$\AA\ %
is the lattice constant. It is instructive to write down the Hamiltonian $%
H_{+}$\ explicitly as%
\begin{equation}
\left( 
\begin{array}{cccc}
-\lambda_{\text{SO}}+\ell E_{z} & v_{\text{F}}k_{+} & ia\lambda_{\text{R}%
}k_{-} & 0 \\ 
v_{\text{F}}k_{-} & \lambda_{\text{SO}}-\ell E_{z} & 0 & -ia\lambda_{\text{R}%
}k_{-} \\ 
-ia\lambda_{\text{R}}k_{+} & 0 & \lambda_{\text{SO}}+\ell E_{z} & v_{\text{F}%
}k_{+} \\ 
0 & ia\lambda_{\text{R}}k_{+} & v_{\text{F}}k_{-} & -\lambda_{\text{SO}%
}-\ell E_{z}%
\end{array}
\right)
\end{equation}
in the basis $\left\{ \psi_{A\uparrow},\psi_{B\uparrow},\psi_{A\downarrow
},\psi_{B\downarrow}\right\} ^{t}$, where $k_{\pm}=k_{x}\pm i k_{y}$. The
two Hamiltonians $H_{+}$ and $H_{-}$ are related through the time-reversal
operation.

The energy spectrum is readily derived from (\ref{DiracHamil}),%
\begin{equation}
\mathcal{E}_{\eta }=\pm \sqrt{\hbar ^{2}v_{\text{F}}^{2}k^{2}+\left( \ell
E_{z}-\eta s_{z}\sqrt{\lambda _{\text{SO}}^{2}+a^{2}\lambda _{\text{R}%
}^{2}k^{2}}\right) ^{2}},  \label{gapDirac}
\end{equation}%
which yields the result (\ref{gap}) to the gap energy at $k=0$.

\section{Inhomogeneous electric field}

The low-energy Dirac theory allows us to investigate analytically the
properties of the helical zero mode under inhomogeneous electric field. In
so doing we set $\lambda _{\text{R}}=0$ to simplify calculations. This
approximation is justified by the following reasons. First all all, we have
numerically checked that the band structure is rather insensitive to $%
\lambda _{\text{R}}$ based on the tight-binding Hamiltonian (\ref{BasicHamil}%
). Second, $\lambda _{\text{R}}$ appears only in the combination $\lambda _{%
\text{R}}k_{\pm }$ in the Hamiltonian (\ref{DiracHamil}), which vanishes
exactly at the K$_{\pm }$ points. Third, the critical electric field $E_{c}$
is independent of $\lambda _{\text{R}}$ as in (\ref{CritiE}).

\subsection{Inhomogeneous electric field along $x$-axis}

We apply the electric field $E_{z}\left( x\right) $ perpendicularly to a
silicene sheet homogeneously in the $y$ direction and inhomogeneously in the 
$x$ direction. We may set $k_{y}=$constant due to the translational
invariance along the $y$ axis. The momentum $k_{y}$ is a good quantum
number. Setting%
\begin{equation}
\Psi \left( x,y\right) =e^{ik_{y}y}\Phi \left( x\right) ,
\end{equation}%
we seek the zero-energy solution, where $\Psi \left( x,y\right) $ is a
four-component amplitude. The particle-hole symmetry guarantees the
existence of zero-energy solutions satisfying the relation $\phi _{B}\left(
x\right) =i\xi \phi _{A}\left( x\right) $ with $\xi =\pm 1$. Here, $\phi
_{A} $ is a two-component amplitude with the up spin and the down spin. Then
the eigenvalue problem yields%
\begin{equation}
H_{\eta }\phi _{A}(x)=E_{\eta \xi }\phi _{A}(x),
\end{equation}%
together with a linear dispersion relation%
\begin{equation}
E_{\eta \xi }=\eta \xi \hbar v_{\text{F}}k_{y}.  \label{DispeRelat}
\end{equation}%
The equation of motion for $\phi _{A}(x)$ reads%
\begin{equation}
\left( \xi \hbar v_{\text{F}}\partial _{x}+\eta \lambda _{\text{SO}}\sigma
_{z}-\ell E_{z}\left( x\right) \right) \phi _{A}(x)=0.
\end{equation}%
We can explicitly solve this as%
\begin{equation}
\phi _{As_{z}}\left( x\right) =C\exp f(x),
\end{equation}%
with 
\begin{equation}
f\left( x\right) =C\exp \left[ \frac{\xi }{\hbar v_{\text{F}}}\int^{x}\left(
-\eta s_{z}\lambda _{\text{SO}}+\ell E_{z}\left( x^{\prime }\right) \right)
dx^{\prime }\right] ,  \label{ZeroModeSolut}
\end{equation}%
where $C$ is the normalization constant. The sign $\xi $ is determined so as
to make the wave function finite in the limit $\left\vert x\right\vert
\rightarrow \infty $. The current is calculated as%
\begin{equation}
J_{s_{z}}\left( x\right) =\text{Re}\left[ \frac{\hbar }{2mi}\Psi
_{s_{z}}^{\dagger }\partial _{y}\Psi _{s_{z}}\right] =\frac{\hbar k_{y}}{m}%
|\phi _{As_{z}}\left( x\right) |^{2}.
\end{equation}%
This is a reminiscence of the Jackiw-Rebbi mode\cite{Jakiw} proposed for the
chiral mode.

The difference between the chiral and helical modes is the presence of the
spin factor $s_{z}$\ in the wave function. As we shall see explicitly in
some examples in what follows, we find the condition either $\eta s_{z}=1$\
or $\eta s_{z}=-1$\ for convergence of the wave function. The condition $%
\eta s_{z}=1$\ implies that the spin is up ($s_{z}=1$) at the K point ($\eta
=1$) and that the spin is down ($s_{z}=-1$) at the K' point ($\eta =-1$).
Consequently, the up-spin electrons flow into the positive $x$-direction
while the down-spin electrons flow into the negative $x$-direction, implying
that the pure spin current flows into the positive $x$-direction. On the
other hand, the condition $\eta s_{z}=-1$\ implies that the pure spin
current flows into the negative $x$-direction.

\begin{figure}[t]
\centerline{\includegraphics[width=0.5\textwidth]{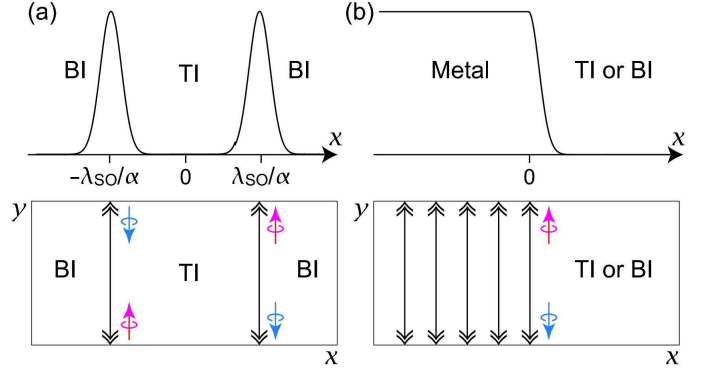}}
\caption{(Color online) The probability density of the helical zero mode
under the electric field $E_{z}$. (a) $E_{z}$ is given by (\protect\ref{EzA}%
) with $\protect\alpha >0$. The spin current flows between a topological
insulator and a band insulator. (b) $E_{z}$ is given by (\protect\ref{EzB})
with $\protect\alpha >0$. The spin current flows in the metallic region. The
arrows indicate the helical mode with spin up at the K point and with spin
down at the K' point. }
\label{FigLocal}
\end{figure}

\textit{Interface between topological and band insulators: }We apply an
electric field such that%
\begin{equation}
E_{z}\left( x\right) =\alpha x/\ell .  \label{EzA}
\end{equation}%
Substituting it to (\ref{ZeroModeSolut}), we obtain%
\begin{equation}
f\left( x\right) =\frac{\xi \alpha }{2\hbar v_{\text{F}}}\left\{ \left( x-%
\frac{\eta s_{z}\lambda _{\text{SO}}}{\alpha }\right) ^{2}-\left( \frac{\eta
s_{z}\lambda _{\text{SO}}}{\alpha }\right) ^{2}\right\} ,
\end{equation}%
where $\xi $\ is chosen to make $\xi \alpha <0$\ for convergence. The wave
function is localized along the two lines $x=\eta s_{z}\lambda _{\text{SO}%
}/\alpha =\pm \lambda _{\text{SO}}/\alpha $\ taken in the bulk, where the
gapless mode emerges since $E_{z}\left( x\right) =\pm \lambda _{\text{SO}}$.
The currents are helical along these two lines, where $\eta s_{z}=1$\ along
one line and $\eta s_{z}=-1$\ along the other line: The spin currents flow
in the opposite directions along these two lines, sandwiched by a band
insulator ($|x|>\lambda _{\text{SO}}/|\alpha |$) and a topological insulator
($|x|<\lambda _{\text{SO}}/|\alpha |$). We illustrate the probability
density $|\phi _{As_{z}}\left( x\right) |^{2}$\ in Fig.\ref{FigLocal}(a) in
the case $\alpha >0$. Each line may be used as a quantum wire. 

\textit{Interface between metal and insulator: }We apply an electric field
such that

\begin{equation}
E_{z}\left( x\right) =E_{c}+\frac{\alpha x}{\ell }\Theta \left( x\right) ,
\label{EzB}
\end{equation}%
where $\Theta \left( x\right) $ is the step function: $\Theta \left(
x\right) =0$ for $x<0$ and $\Theta \left( x\right) =1$ for $x\geq 0$.
Substituting it to (\ref{ZeroModeSolut}), we obtain%
\begin{equation}
f\left( x\right) =\frac{\xi }{\hbar v_{\text{F}}}\left\{ \frac{\alpha x^{2}}{%
2}\Theta \left( x\right) +\left( 1-\eta s_{z}\right) \lambda _{\text{SO}%
}x\right\} ,
\end{equation}%
where $\xi $\ is chosen to make $\xi \alpha <0$\ for convergence. Namely,
when we choose $\alpha >0$, we find%
\begin{equation}
\begin{array}{ll}
E_{z}\left( x\right) =E_{c}\quad \text{(metal)} & \text{for}\quad x\leq 0 \\ 
E_{z}\left( x\right) >E_{c}\quad \text{(BI)} & \text{for}\quad x>0%
\end{array}%
,
\end{equation}%
and when we choose $\alpha <0$, we find%
\begin{equation}
\begin{array}{ll}
E_{z}\left( x\right) =E_{c}\quad \text{(metal)} & \text{for}\quad x\leq 0 \\ 
E_{z}\left( x\right) <E_{c}\quad \text{(TI)} & \text{for}\quad x>0%
\end{array}%
,
\end{equation}%
where TI and BI stand for topological and band insulators. In the both cases
the region $x\leq 0$ is a metal. Furthermore, it is necessary that $\eta
s_{z}=1$ for convergence of the wave function, as implies that the current
is helical in the metallic region. We illustrate the probability density $%
|\phi _{A\uparrow }\left( x\right) |^{2}$ in Fig.\ref{FigLocal}(b) in the
case $\alpha >0$, which is a constant for $x\leq 0$.

It is intriguing that there emerge helical zero modes in metal. This is not
surprising because the edge of a topological insulator is a sufficient
condition but not a necessary condition for the emergence of helical zero
modes. The helical zero mode requires the massless Dirac fermion, the
time-reversal symmetry and the spin-orbit interaction. Our example shows
explicitly that it can appear in regions besides the edge of a topological
insulator provided these conditions are satisfied.

\begin{figure}[t]
\centerline{\includegraphics[width=0.45\textwidth]{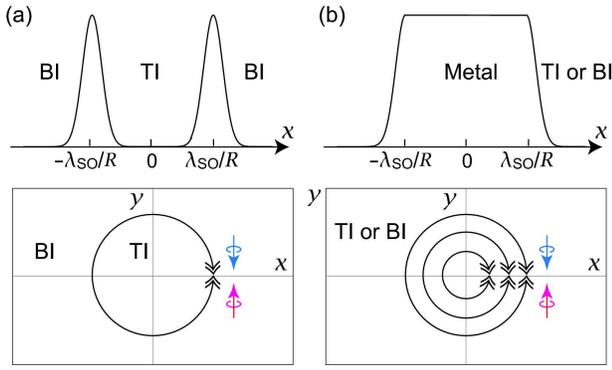}}
\caption{(Color online) The probability density of the helical zero mode
under the electric field $E_{z}$ with $\protect\alpha >0$. (a) $E_{z}$ is
given by (\protect\ref{EzD}). The spin current flows between a topological
insulator and a band insulator by encircling the topological insulator. (b) $%
E_{z}$ is given by (\protect\ref{EzC}) with $\protect\alpha >0$. The spin
current flows in the disklike metallic region confined within a topological
or band insulator. The arrows indicate the helical mode with spin up at the
K point and with spin down at the K' point. }
\label{FigPond}
\end{figure}

\subsection{Inhomogeneous electric field along $r$-axis}

We apply a cylindrical symmetric inhomogeneous electric field $E_{z}\left(
r\right) $ to a silicene sheet. The equation reads%
\begin{equation}
\left( 
\begin{array}{cc}
-\lambda_{\text{SO}}\sigma_{z}+\ell E_{z}\left( r\right) & \hbar v_{\text{F}%
}e^{i\theta}\left( i\partial_{r}-\frac{1}{r}\partial_{\theta }\right) \\ 
\hbar v_{\text{F}}e^{-i\theta}\left( i\partial_{r}+\frac{1}{r}\partial
_{\theta}\right) & \lambda_{\text{SO}}\sigma_{z}-\ell E_{z}\left( r\right)%
\end{array}
\right) \left( 
\begin{array}{c}
\psi_{A} \\ 
\psi_{B}%
\end{array}
\right) =0.
\end{equation}
We solve this for zero-energy states by setting 
\begin{equation}
\left( 
\begin{array}{c}
\psi_{A}(r,\theta) \\ 
\psi_{B}(r,\theta)%
\end{array}
\right) =\left( 
\begin{array}{c}
e^{i\eta\theta/2}\phi_{A}\left( r\right) \\ 
e^{-i\eta\theta/2}\phi_{B}\left( r\right)%
\end{array}
\right) .
\end{equation}
The equation of motion $H_{K}\psi=0$ is transformed into 
\begin{equation}
\left( \xi\hbar v_{\text{F}}\left( \partial_{r}+\frac{1}{2r}\right)
+\eta\lambda_{\text{SO}}\sigma_{z}-\ell E_{z}\left( r\right) \right)
\phi_{A}=0,
\end{equation}
which we solve as%
\begin{equation}
\phi_{As_{z}}\left( r\right) =\frac{C}{\sqrt{r}}\exp f(r),
\end{equation}
with 
\begin{equation}
f\left( r\right) =\frac{\xi}{\hbar v_{\text{F}}}\int_{0}^{r}\left( -\eta
s_{z}\lambda_{\text{SO}}+\ell E_{z}\left( r^{\prime}\right) \right)
dr^{\prime},  \label{RadiaZeroMode}
\end{equation}
where $C$ is the normalization constant and $\xi=\pm1$. The sign $\xi$ is
determined so as to make the wave function finite in the limit $r\rightarrow
\infty$.

\textit{Interface between topological and band insulators: }We apply an
electric field such that%
\begin{equation}
E_{z}\left( r\right) =\alpha r/\ell .  \label{EzD}
\end{equation}%
Substituting it to (\ref{RadiaZeroMode}), we have%
\begin{equation}
f\left( r\right) =\frac{\xi \alpha }{2\hbar v_{\text{F}}}\left\{ \left( r-%
\frac{\eta s_{z}\lambda _{\text{SO}}}{\alpha }\right) ^{2}-\left( \frac{\eta
s_{z}\lambda _{\text{SO}}}{\alpha }\right) ^{2}\right\} .
\end{equation}%
where $\xi $\ is chosen to make $\xi \alpha <0$\ for convergence. The wave
function is localized along the circle $r=\eta s_{z}\lambda _{\text{SO}%
}/\alpha >0$, where $E_{z}\left( r\right) =\pm \lambda _{\text{SO}}$. When
we choose $\alpha >0$\ it is necessary that $\eta s_{z}=1$, and when we
choose $\alpha <0$\ it is necessary that $\eta s_{z}=-1$. In any of the two
cases, there emerges helical zero modes and the spin current flows along the
circle between a topological insulator ($r<\lambda _{\text{SO}}/|\alpha |$)
and a band insulator ($r>\lambda _{\text{SO}}/|\alpha |$). The direction of
the spin current is opposite for $\alpha >0$\ and $\alpha <0$. We illustrate
the probability density $|\phi _{As_{z}}\left( r\right) |^{2}$\ in Fig.\ref%
{FigPond}(a) in the case $\alpha >0$.  This region may be used as a quantum
wire.

\textit{Interface between metal and insulator: }We apply an electric field
such that%
\begin{equation}
E_{z}\left( r\right) =E_{c}+\frac{\alpha (r-R)}{\ell }\Theta \left(
r-R\right) ,  \label{EzC}
\end{equation}%
where $\Theta \left( r\right) $ is the step function. Substituting it to (%
\ref{RadiaZeroMode}), we have%
\begin{equation}
f\left( r\right) =\frac{\xi }{\hbar v_{\text{F}}}\left\{ \frac{\alpha
(r-R)^{2}}{2\ell }\Theta \left( r-R\right) +\left( 1-\eta s_{z}\right)
\lambda _{\text{SO}}(r-R)\right\} ,
\end{equation}%
where $\xi $\ is chosen to make $\xi \alpha <0$\ for convergence. It is
notable that $E_{z}\left( r\right) =E_{c}$ for $r<R$ and hence the system is
metallic there. The wave function describes an interface between a metal for 
$r<R$ and an insulator for $r>R$. The insulator is a topological insulator
when we choose $\alpha <0$ and a band insulator when we choose $\alpha >0$.
Since it is necessary that $\eta s_{z}=1$ for convergence of the wave
function, the current is helical in the metallic region ($r\leq R$). We show
the probability density $|\phi _{A\uparrow }\left( r\right) |^{2}$ in Fig.%
\ref{FigPond}(b), where $\phi _{A\uparrow }\left( r\right) =$constant for $%
r\leq R$. This region may act as a quantum dot.

\section{Conclusions}

\label{SecConclusion}

Taking advantage of the buckled structure of silicene we have demonstrated
that we can control its band structure by applying the electric field $E_{z}$%
. Silicene undergoes a topological phase transition between a topological
insulator and a band insulator as $E_{z}$ crosses the critical point $\pm
E_{c}$. It is a semimetal at $E_{z}=\pm E_{c}$.

A novel phenomenon appears when we apply an inhomogeneous electric field. We
have explicitly constructed wave functions of helical zero modes for simple
geometrical regions based on the low-energy effective Dirac theory. The
results imply in general that helical zero modes can be confined in any
regions by tuning the external electric field locally to the critical field (%
\ref{CritiE}), $E_{z}(x,y)=E_{c}$. Our system may be the first example in
which helical zero modes appear in regions besides the edge of a topological
insulator. It is to be emphasized that we can apply an inhomogeneous
electric field so that a single silicene sheet contains several regions
which are topological insulators, band insulators and metals. Such a
structure may open a way for future spintronics. Our results are also
applicable to germanene, that is a two-dimensional honeycomb structure made
of germanium.

I am very much grateful to N. Nagaosa for many fruitful discussions on the
subject. This work was supported in part by Grants-in-Aid for Scientific
Research from the Ministry of Education, Science, Sports and Culture No.
22740196.

\end{document}